\documentclass[aps,prl,twocolumn,showpacs,superscriptaddress,floatfix]{revtex4-1}
\usepackage{bm}
\usepackage{amssymb}
\usepackage{graphicx}
\usepackage{amsmath}
\usepackage{dcolumn}
\usepackage{bm}
\usepackage{color}
\usepackage[german,english]{babel}

\usepackage{array}

\newcolumntype{C}[1]{>{\centering\arraybackslash}p{#1}}
\usepackage{epstopdf}

\def\abinitio{\textit{ab initio}}
\def\G0W0{G$_0$W$_0$}

\begin{document}

\title{Electronic Structures and Optical Properties of Partially and Fully Fluorinated Graphene}

\author{Shengjun Yuan}
\email{s.yuan@science.ru.nl}
\affiliation{Institute for Molecules and Materials, Radboud
University Nijmegen, Heyendaalseweg 135, 6525AJ Nijmegen, The Netherlands}
\author{Malte R\"osner}
\affiliation{Institut f{\"u}r Theoretische Physik, Universit{\"a}t Bremen, Otto-Hahn-Allee 1, 28359 Bremen, Germany}
\affiliation{Bremen Center for Computational Materials Science, Universit{\"a}t Bremen, Am Fallturm 1a, 28359 Bremen, Germany}
\author{Alexander Schulz}
\affiliation{Institut f{\"u}r Theoretische Physik, Universit{\"a}t Bremen, Otto-Hahn-Allee 1, 28359 Bremen, Germany}
\affiliation{Bremen Center for Computational Materials Science, Universit{\"a}t Bremen, Am Fallturm 1a, 28359 Bremen, Germany}
\author{Tim O. Wehling}
\affiliation{Institut f{\"u}r Theoretische Physik, Universit{\"a}t Bremen, Otto-Hahn-Allee 1, 28359 Bremen, Germany}
\affiliation{Bremen Center for Computational Materials Science, Universit{\"a}t Bremen, Am Fallturm 1a, 28359 Bremen, Germany}
\author{Mikhail I. Katsnelson}
\affiliation{Institute for Molecules and Materials, Radboud
University Nijmegen, Heyendaalseweg 135, 6525AJ Nijmegen, The Netherlands}

\date{\today}

\begin{abstract}
In this letter we study the electronic structures and optical properties of partially and fully fluorinated graphene by a combination of \abinitio\ \G0W0\ calculations and large-scale multi-orbital tight-binding simulations. We find that for partially fluorinated graphene, the appearance of paired fluorine atoms is more favorable than unpaired atoms. We also show that different types of structural disorder, such as carbon vacancies, fluorine vacancies, fluorine vacancy-clusters and fluorine armchair- and zigzag-clusters, will introduce different types of midgap states and extra excitations within the optical gap. Furthermore we argue that the local formation of $sp^3$ bonds upon fluorination can be distinguished from other disorder inducing mechanisms which do not destroy the $sp^2$ hybrid orbitals by measuring the polarization rotation of passing polarized light.


\end{abstract}


\maketitle

Fluorinated graphene has attracted great interest over the past few years\cite{Nair2010, Zboril2010, Withers2010, Cheng2010, Robinson2010, Nair2012}. Unlike graphene which is a two-dimensional semiconductor with zero energy band gap, fully fluorinated graphene (or fluorographene, graphene fluoride) is a wide gap semiconductor. The experimentally observed optical band gaps vary between $3\,$eV\cite{Nair2010} and $3.8\,$eV\cite{Jeon2011}, which is comparable to the result of standard density functional theory (DFT) calculations\cite{Karlicky2013}. However, the tendency of DFT calculations to underestimate band gaps is well known. Thus, high-level many-body calculations in the GW approximations have been used to calculate the band gap as well. These yield a quasiparticle band gap $E_{gap}\approx 7\,$eV, which is approximately twice larger than the experimentally observed optical excitation gaps\cite{KLKE10, Leenaerts2010, Samarakoon2011, Karlicky2013}. A natural candidate to explain this discrepancy of 
calculated quasiparticle gaps and optical experiments are excitonic effects, which were considered in recent calculations of optical spectra based on the Bethe-Saltpeter equation (BSE). These yield an optical band gap of $5.1\,$eV \cite{Karlicky2013}, which means sizable excitonic effects but which is still larger than the experimental values. It is argued but not verified that the remaining gap between the theoretical calculations and the experimental observers might be the result of disorder introduced during the fluorination process. To clarify this issue, we perform a systematic study of different types of structural disorder by a combination of \abinitio\ calculations and large-scale tight-binding (TB) simulations.

\begin{figure*}[t]
  \begin{center}
    \mbox{
      \includegraphics[width=7.5cm]{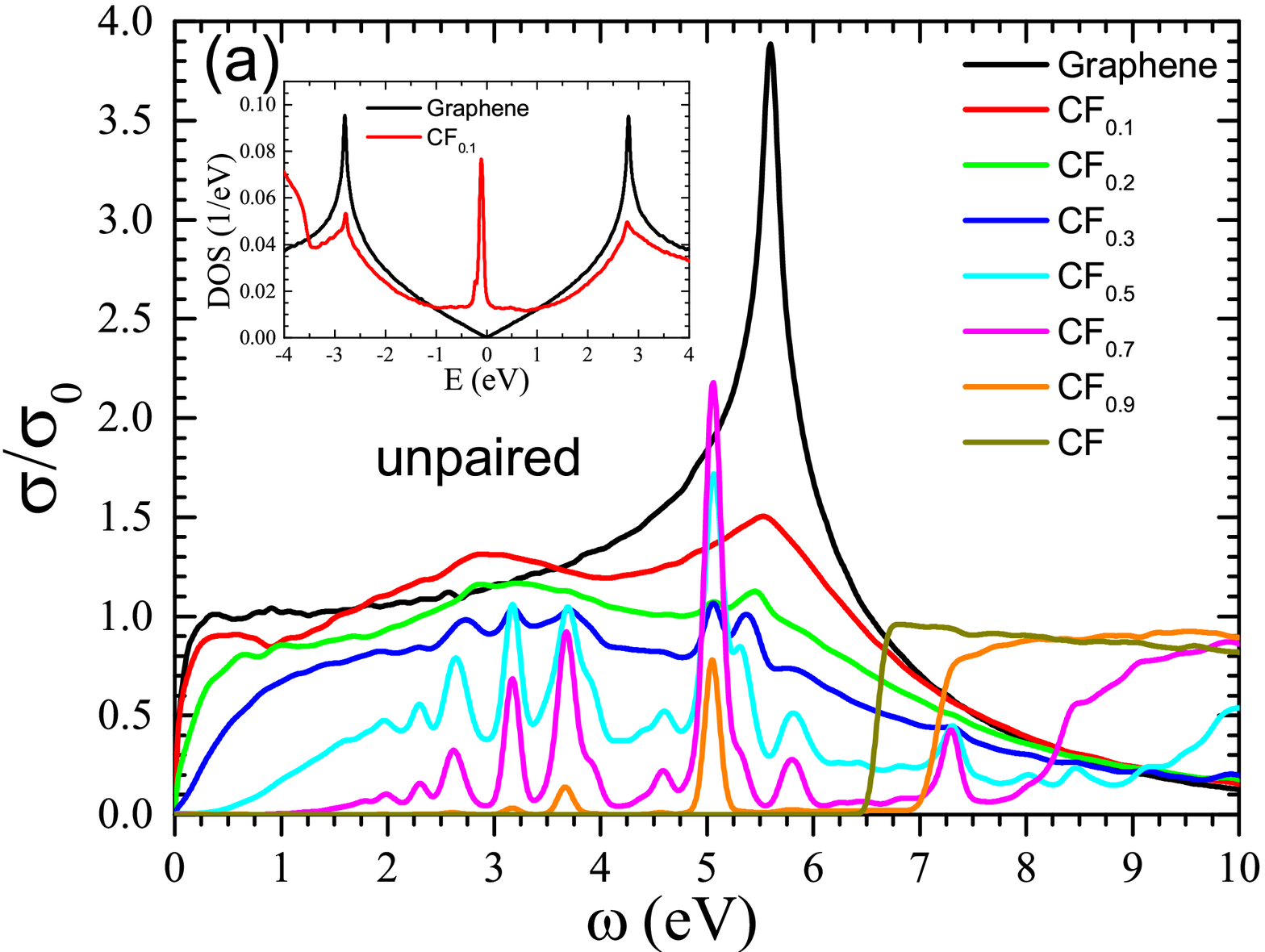}
      \includegraphics[width=7.5cm]{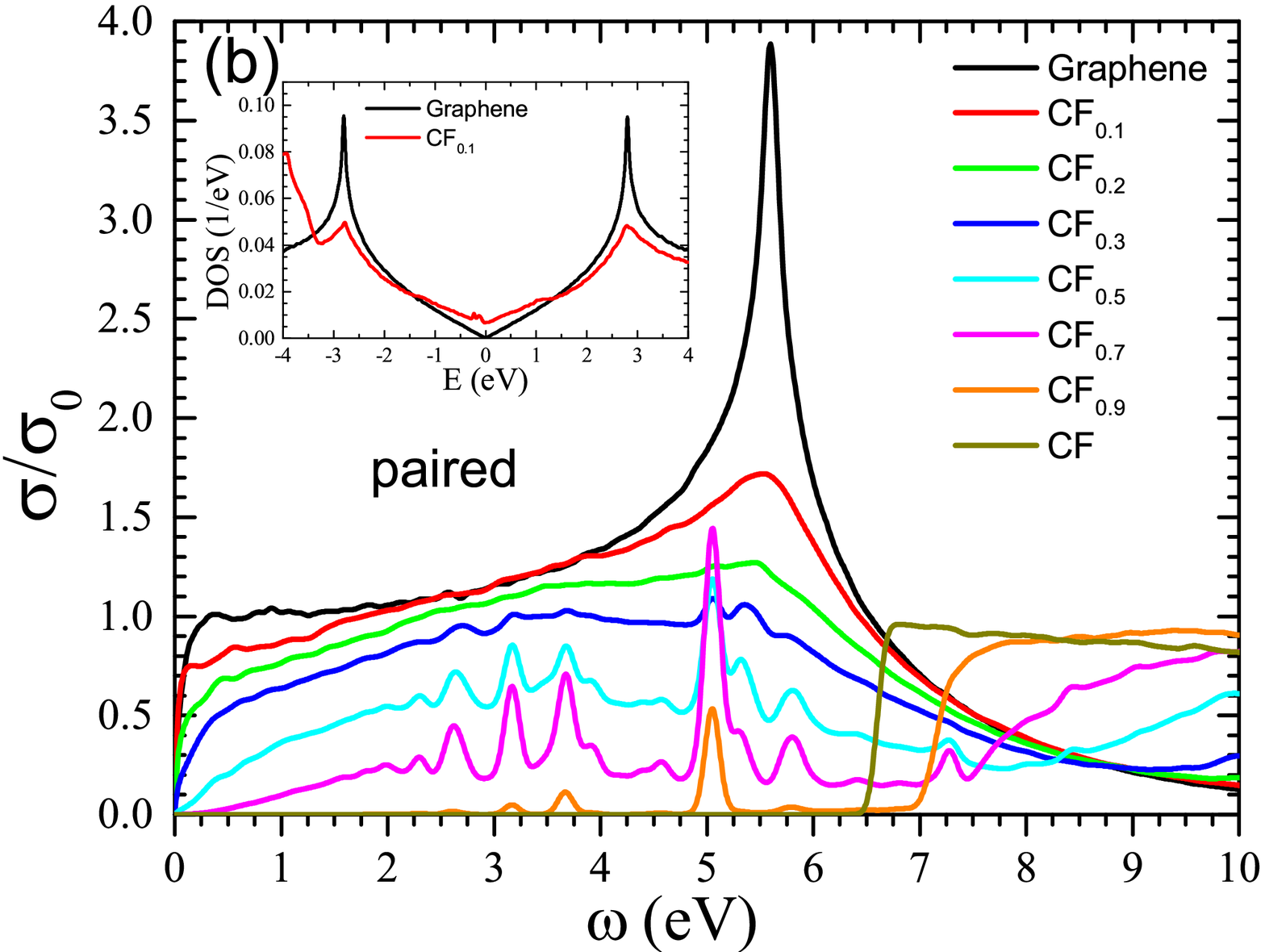}
    }
    \mbox{
      \includegraphics[width=7.5cm]{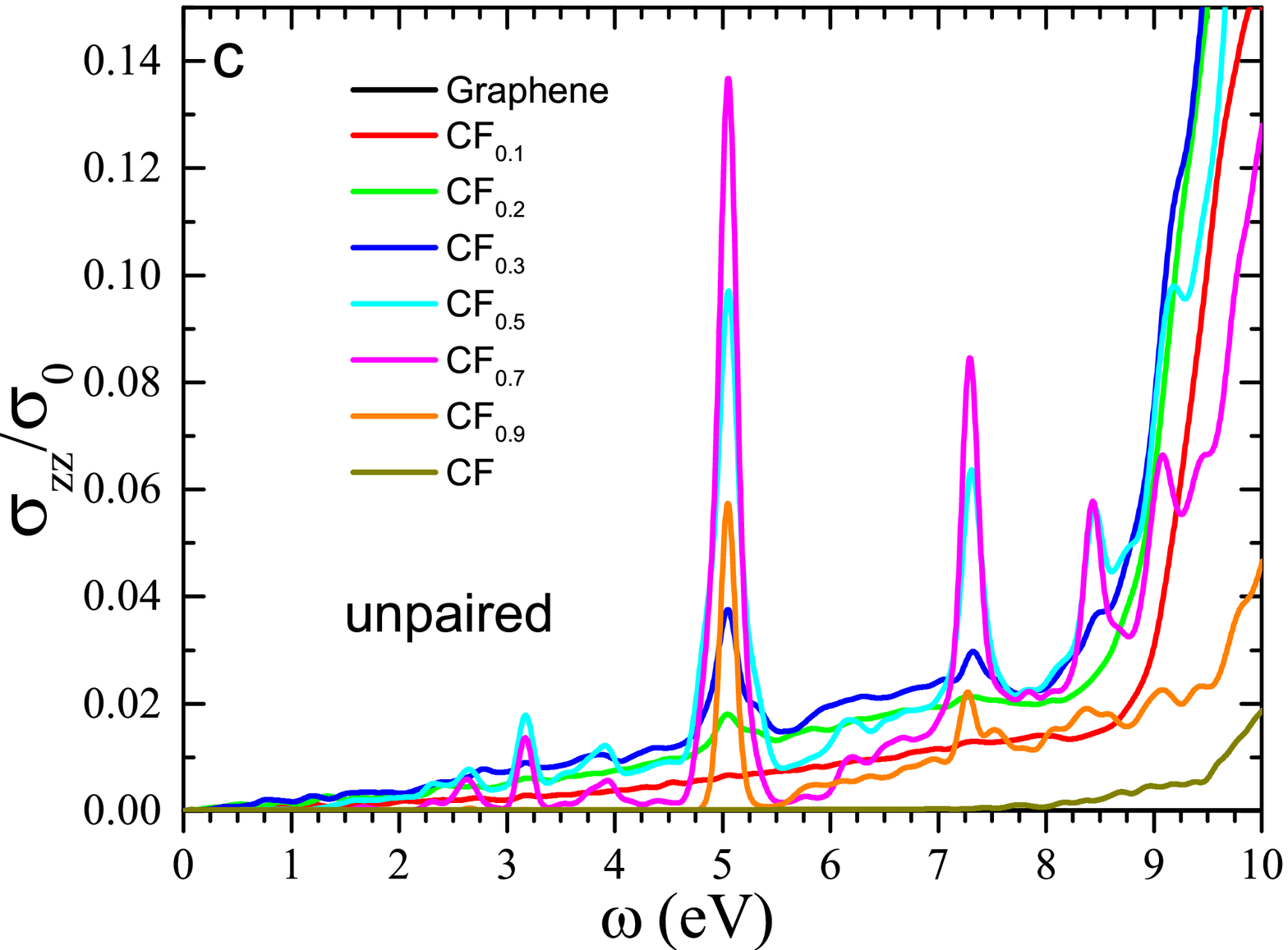}
      \includegraphics[width=7.5cm]{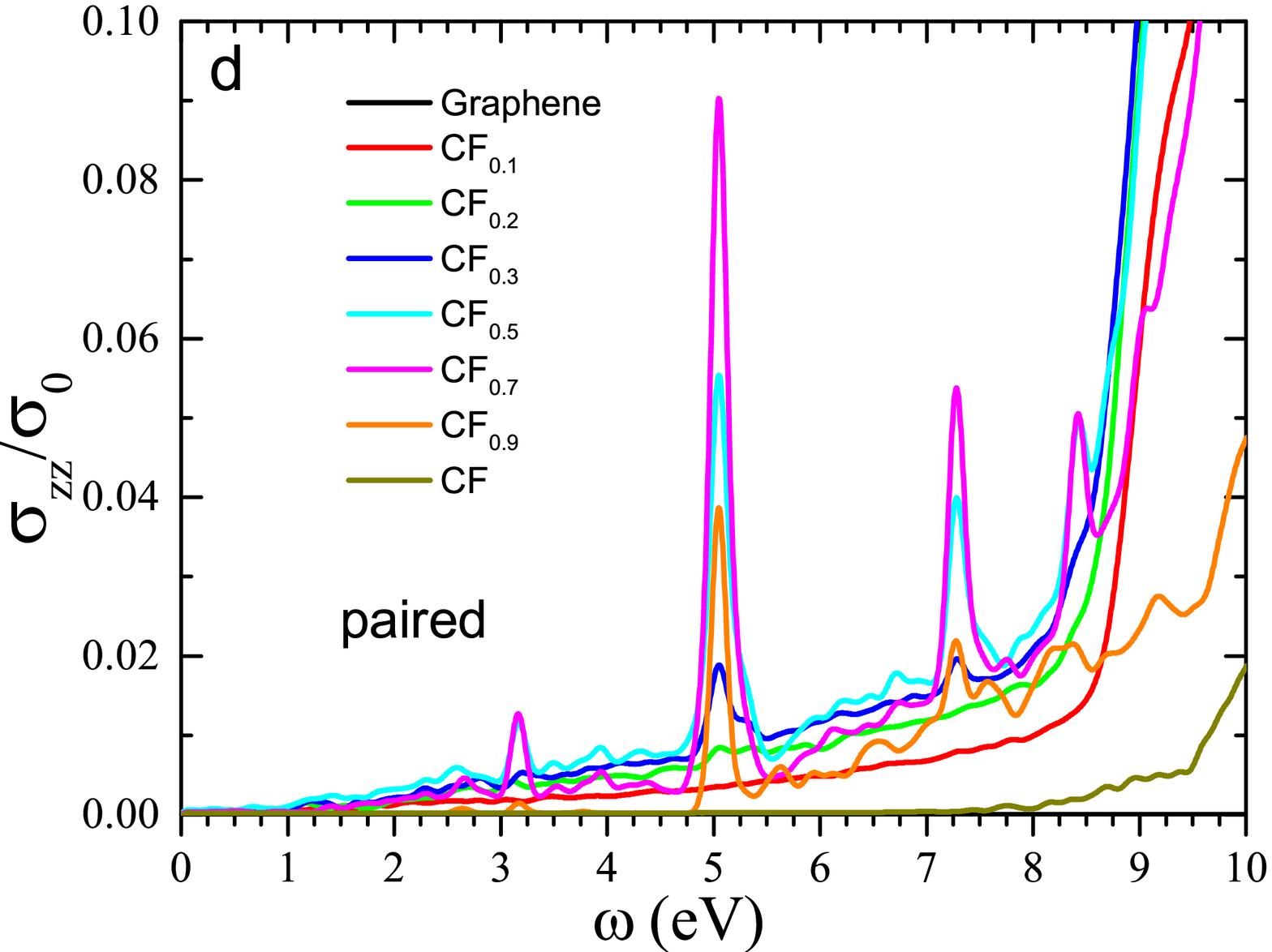}
    }
  \end{center}
  \caption{In-plane (a,b) and out-of-plane (c,d) optical conductivity of partially and fully fluorinated graphene with different concentration of randomly distributed unpaired or paired fluorine adatoms. The density of states of graphene and CF$_{0.1}$ are plotted as inserts of panels (a,b) and reveal midgap states in the unpaired case (see the sharp peak close to the neutrality point).}
  \label{Fig:ac}
\end{figure*}

Upon fluorination the corresponding carbon atom will move out of the graphene plane and the former $sp^2$ hybrid orbitals (formed by the carbon $s$, $p_x$ and $p_y$ orbitals) will change to a $sp^3$ hybrid orbital (including the $p_z$ orbital) due to the additional bond to the fluorine atom. Thus, an extended TB model describing this $sp^2$/$sp^3$ transition when going from (i) pristine graphene via (ii)partially fluorinated graphene to (iii) pristine fluorographene is required. Therefore, we construct a nearest neighbor TB model accounting for four orbitals ($2s,2p_{x},2p_{y},2p_{z}$) per carbon and three orbitals ($2p_{x},2p_{y},2p_{z}$) per fluorine atom. The TB parameters are derived from  \abinitio\ calculations by fitting
 band structures and local density of states of graphene, partially and fully fluorinated graphene.\footnote{See Suppl. Mat. for more details.} The involved \abinitio\ calculations are carried out in the \G0W0\ approximation. The resulting TB model is by construction very general and capable of describing arbitrary fluorination patterns.

In order to model realistic samples in the TB calculations, we perform simulations of systems on the scale of micrometer, consisting of $2400 \times 2400$ carbon atoms. Thereby we consider for neighboring F atoms exclusively the chair configuration which has been proven to be the most stable structure of fluorinated graphene \cite{Leenaerts2010,Samarakoon2011}. The density of states and optical conductivity are calculated using the TB propagation method \cite{YRK10,WK10}, which is based on the numerical simulation of random wave propagation according to the time-dependent Schr\"odinger equation\footnote{See Suppl. Mat. for more details.}.



 Fig.~\ref{Fig:ac} displays the optical spectrum of fluorinated graphene with different C/F ratios. Here, we consider two types of fluorination: (I) the fluorine atoms are distributed randomly without any correlations between the sublattices and (II) two fluorine atoms are always adsorbed at neighboring carbon atoms. Due to the chair configuration in the latter case the first fluorine atom will be above the carbon plane while the second one will be located beneath the plane. 
It is obvious that in the unpaired case the sublattice symmetry can be locally broken due to different amounts of fluorine atoms on each sublattice. This leads to  midgap states, which can be observed as huge peaks near the neutrality point in the density of states (see inset of Fig.~\ref{Fig:ac} (a)). In turn  additional electron-hole excitations arise in the unpaired case due to transitions between these midgap states and the $\pi$-band saddle-point singularities, which manifest as enhancements of the optical conductivity at energies around $2.8\,$eV for small fluorine concentrations of CF$_{0.1}$, CF$_{0.2}$ and CF$_{0.3}$ in Fig.~\ref{Fig:ac}(a). These enhancements of the optical conductivity around $2.8\,$eV neither appear in the light absorption of partially fluorinated graphene measured in Ref.\cite{Nair2010} nor in the simulated spectra of graphene with paired fluorine adsorbates. Indeed, the experimental absorption spectrum\cite{Nair2010} is close to the one we obtain for CF$_{0.3}$ in the paired fluorination 
case. Altogether this leads us to the conclusion, that the fluorine atoms tend to form pairs during the fluorination process. 

Magnetic measurements for partially fluorinated graphene \cite{nair12} show a small concentration of local spin one-half magnetic moments (roughly, one magnetic moment per thousand of fluorine atoms). Magnetic moments in graphene are associated to mid-gap states \cite{KatsnelsonBook}; thus, our conclusion that the most fluorine atoms form pairs which have no such states and are therefore obviously nonmagnetic, seems to be in agreement with this observation. The residual magnetic moments can be in principle related to the individual fluorine atoms, however, this issue requires further investigation.

Optical experiments can work at normal as well as grazing incidence and measure polarization dependent spectra. We therefore investigate the out-of-plane optical conductivity along the $z$ direction ($\sigma _{zz}$) and compare to the in-plane optical conductivity. The dipole operator associated with $\sigma _{zz}$ contains two parts: one is the electron hopping between the carbon atoms which have different $z$ coordinates, and the other is the hopping between carbon atoms and absorbed fluorine directly  above or below. This results in a zero optical conductivity along the $z$ direction in pristine graphene over the whole spectrum, since there are no differences in the $z$ positions of the carbon atoms. More generally, there are no interatomic contributions to $\sigma_{zz}$ from any $sp^2$-like carbon part of the sample. The evolution of $\sigma _{zz}$ upon random and pair fluorination is shown in Fig.~\ref{Fig:ac} (c) and (d), respectively. Unlike the in-plane optical conductivity, the out-of-plane 
conductivities $\sigma _{zz}$ are similar for both unpaired and paired cases in the energy range shown in  Fig.~\ref{Fig:ac}, independently of the fluorine concentration. Only at higher energies (between $20$ and $25\,$eV) the spectra for the two cases differ noticeably (see Supplementary Materials). There are in particular no features in $\sigma _{zz}$ due to the chiral midgap states associated with local sublattice symmetry breaking in the randomly fluorinated graphene. Thus, polarization analysis of optical spectra yields clear fingerprints for spectral features associated with chiral midgap states. 

Generally, the nonzero optical conductivity perpendicular to the sheets raises the possibility to rotate the polarization of passing polarized light. As nonzero $\sigma_{zz}$ requires the formation of sp$^3$ orbitals, one is able to distinguish between impurity states originating from adatoms to other in-plane disorder configurations (for example, carbon vacancies, in-plane carbon reconstructions like pentagon-heptagon rings, and coulomb impurities) by measuring the polarization angle. 

As can be seen in Fig.~\ref{Fig:ac} (c) and (d), the general trend of $\sigma _{zz}$ at energies below $\sim 10\,$eV is to increase with fluorination up to fluorine concentration of about $30\%$ and to decrease afterwards. Interestingly, there are a few sharp resonances (e.g. around $5$\,eV) in $\sigma _{zz}$ which intensify up to much larger fluorine concentrations on the order of $70\%$. As it will be argued in the following, the peak of the optical conductivity at about $5\,$eV results from fluorine vacancies, which are not well defined for small fluorine concentrations and which will nearly vanish for high concentrations. Thus, this peak arises not before a certain threshold and vanishes towards fully fluorination. For fully fluorinated graphene, $\sigma_{zz}$ becomes zero below the electronic band gap, but is highly enhanced at higher energy (see Supplementary Materials for more details).

\begin{figure*}
  \begin{center}
    \mbox{
      \includegraphics[width=4.5cm]{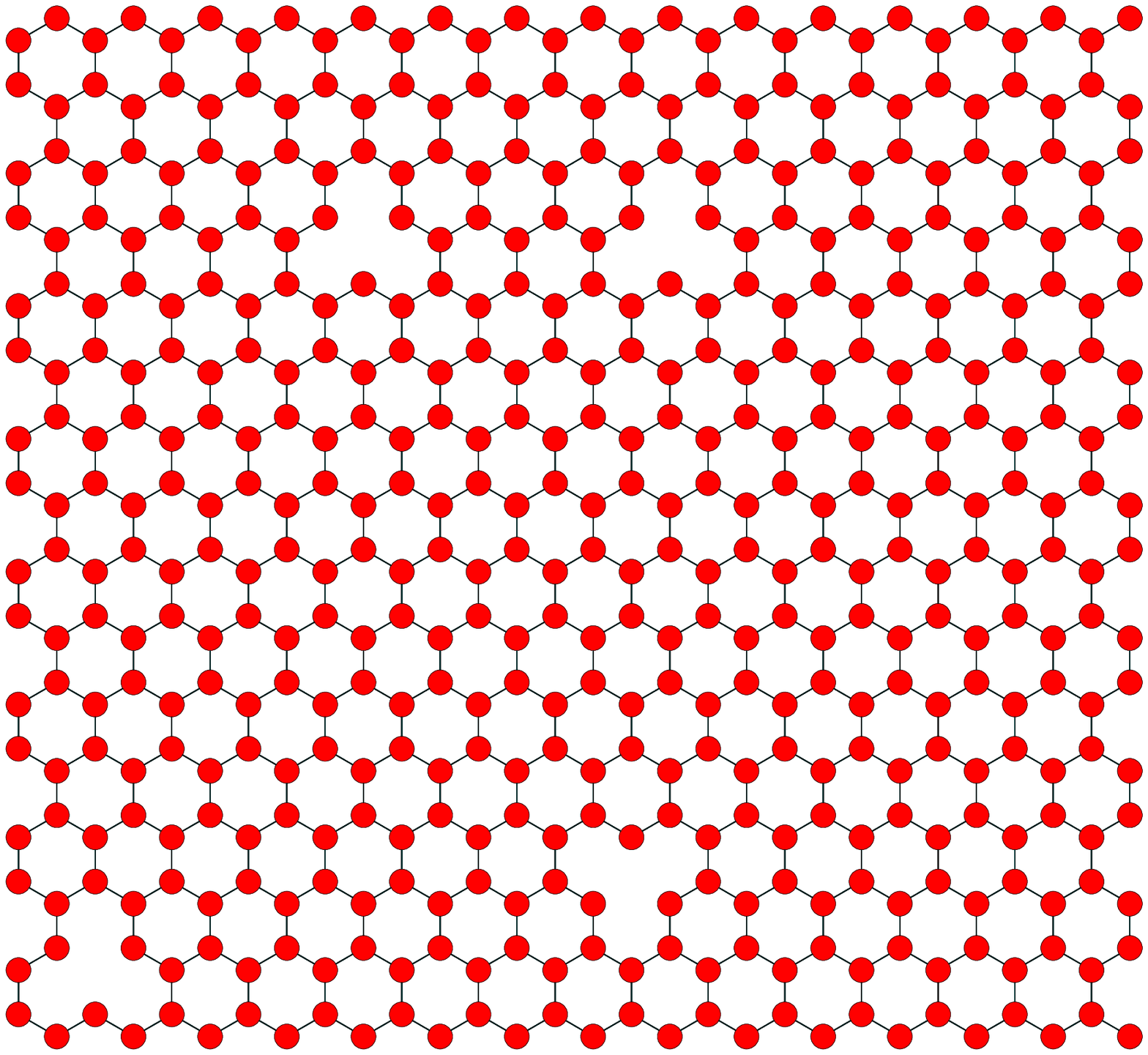}
      \includegraphics[width=5cm]{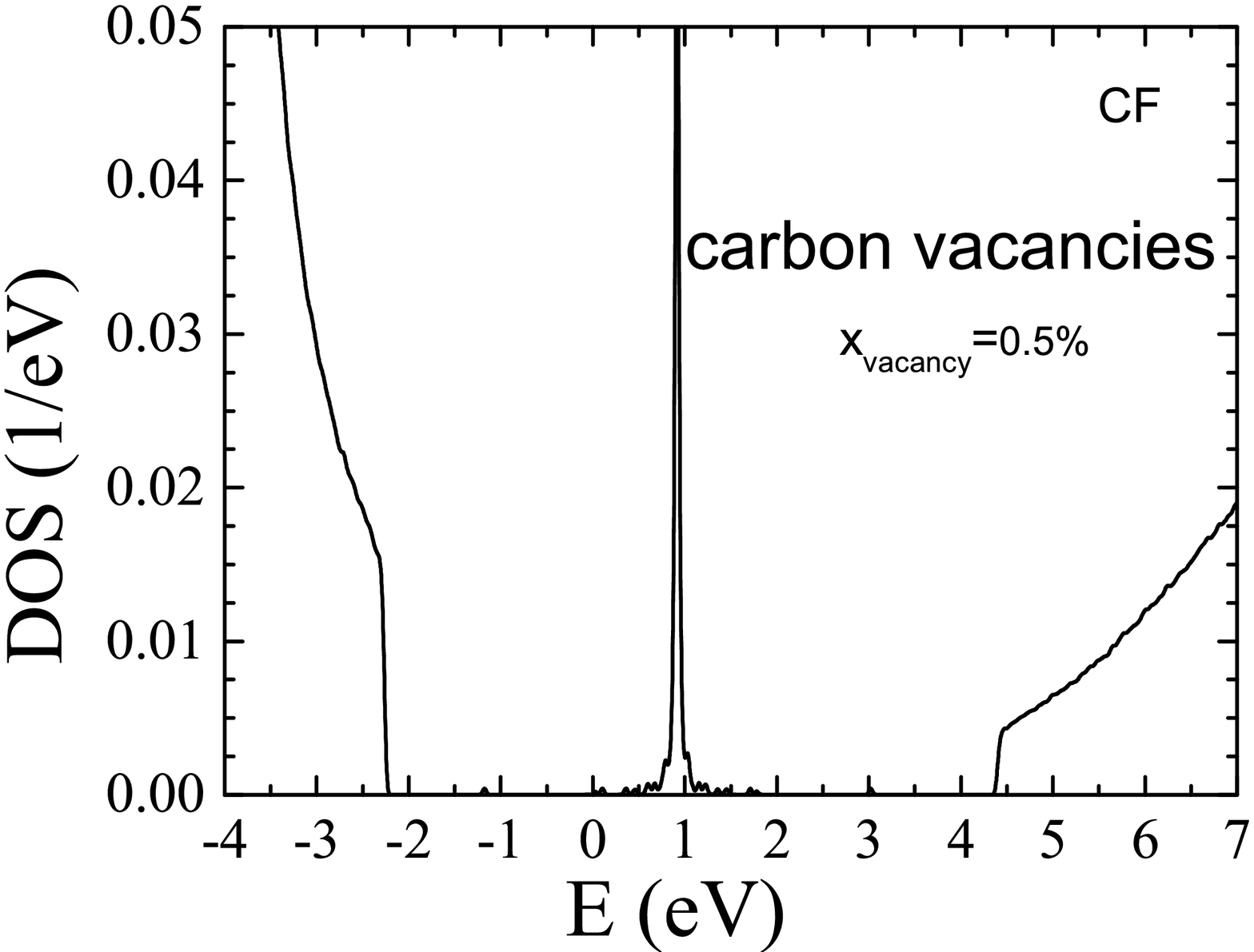}
      \includegraphics[width=5cm]{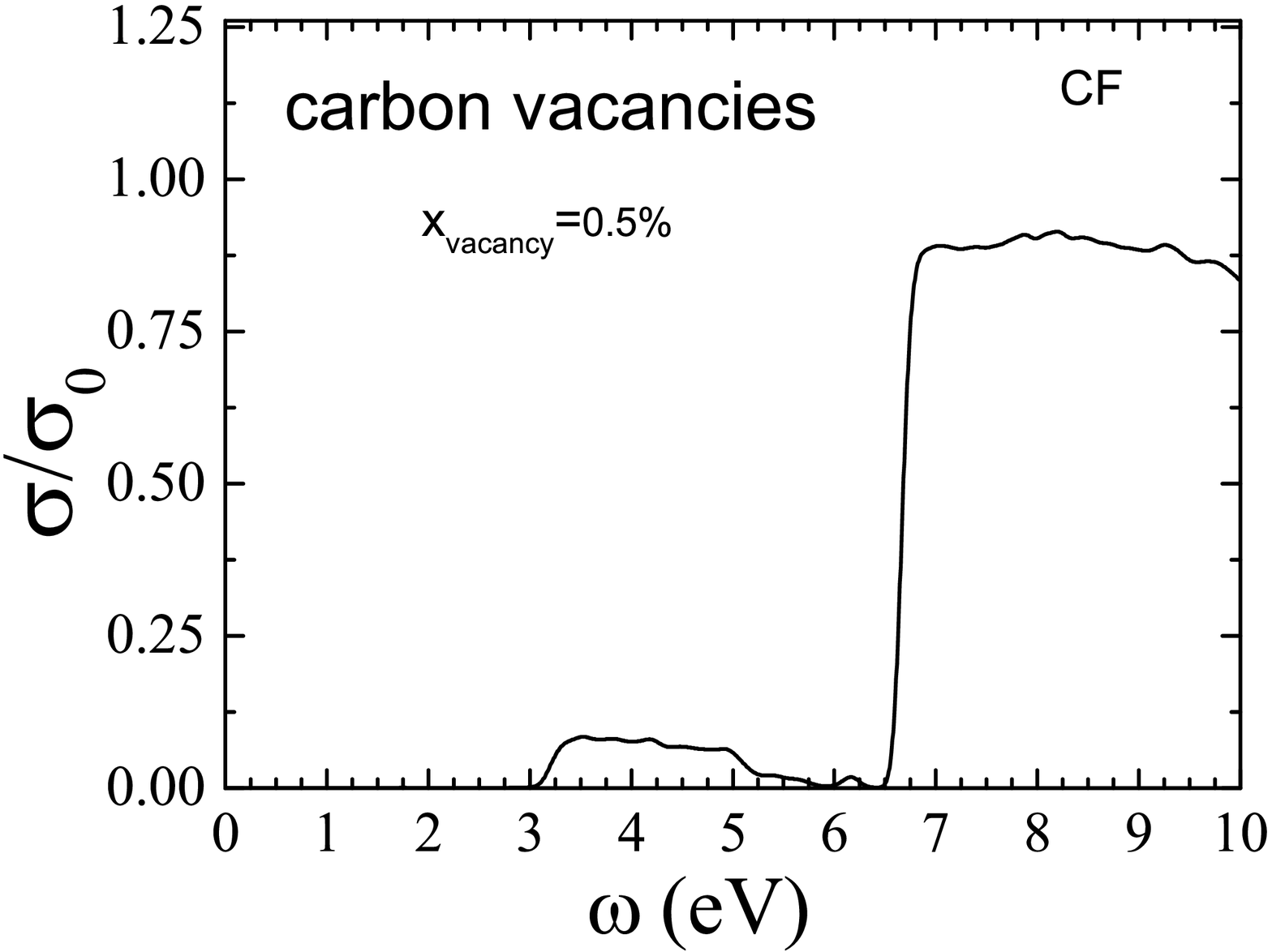}
    }
    \mbox{
      \includegraphics[width=4.5cm]{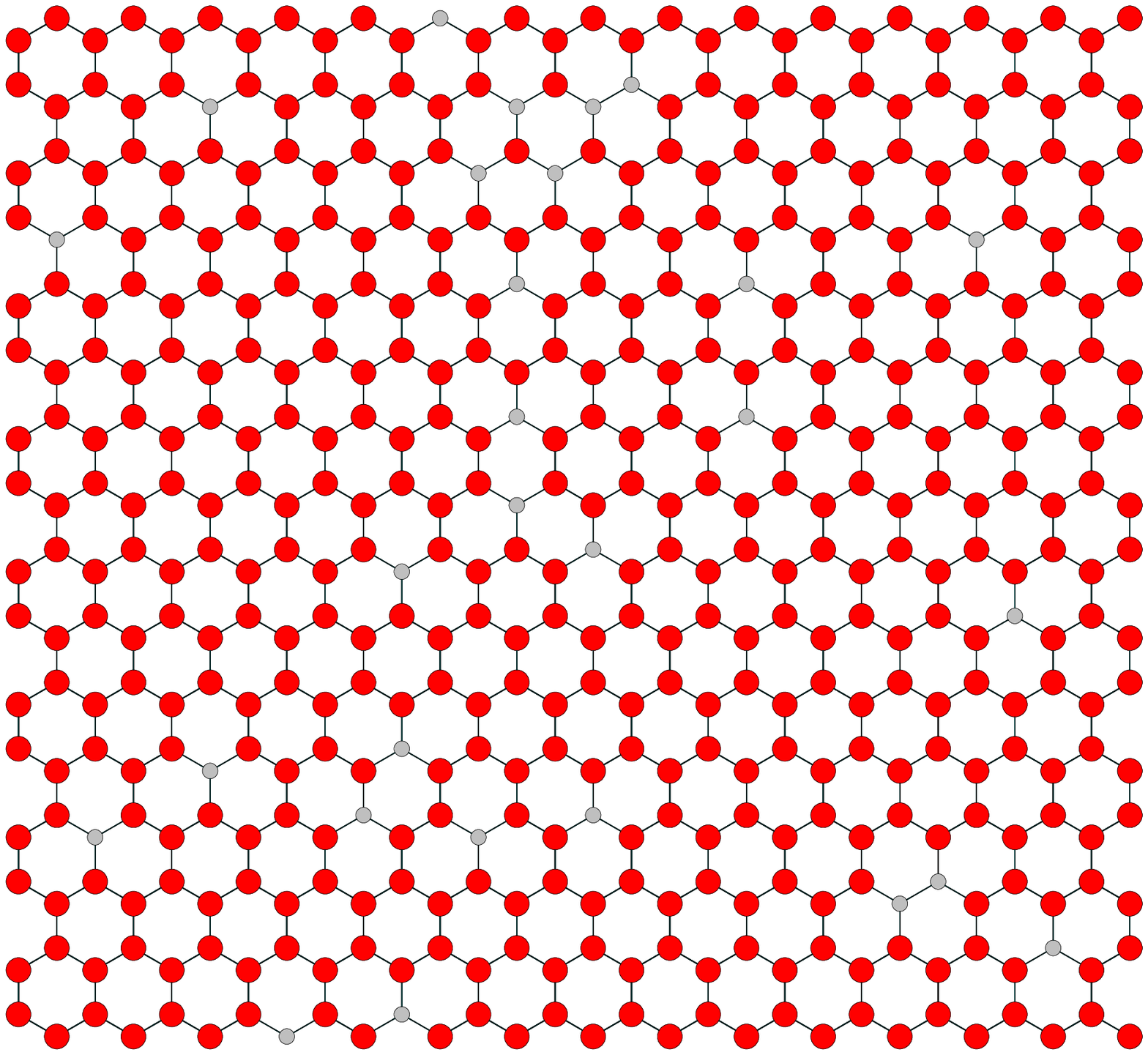}
      \includegraphics[width=5cm]{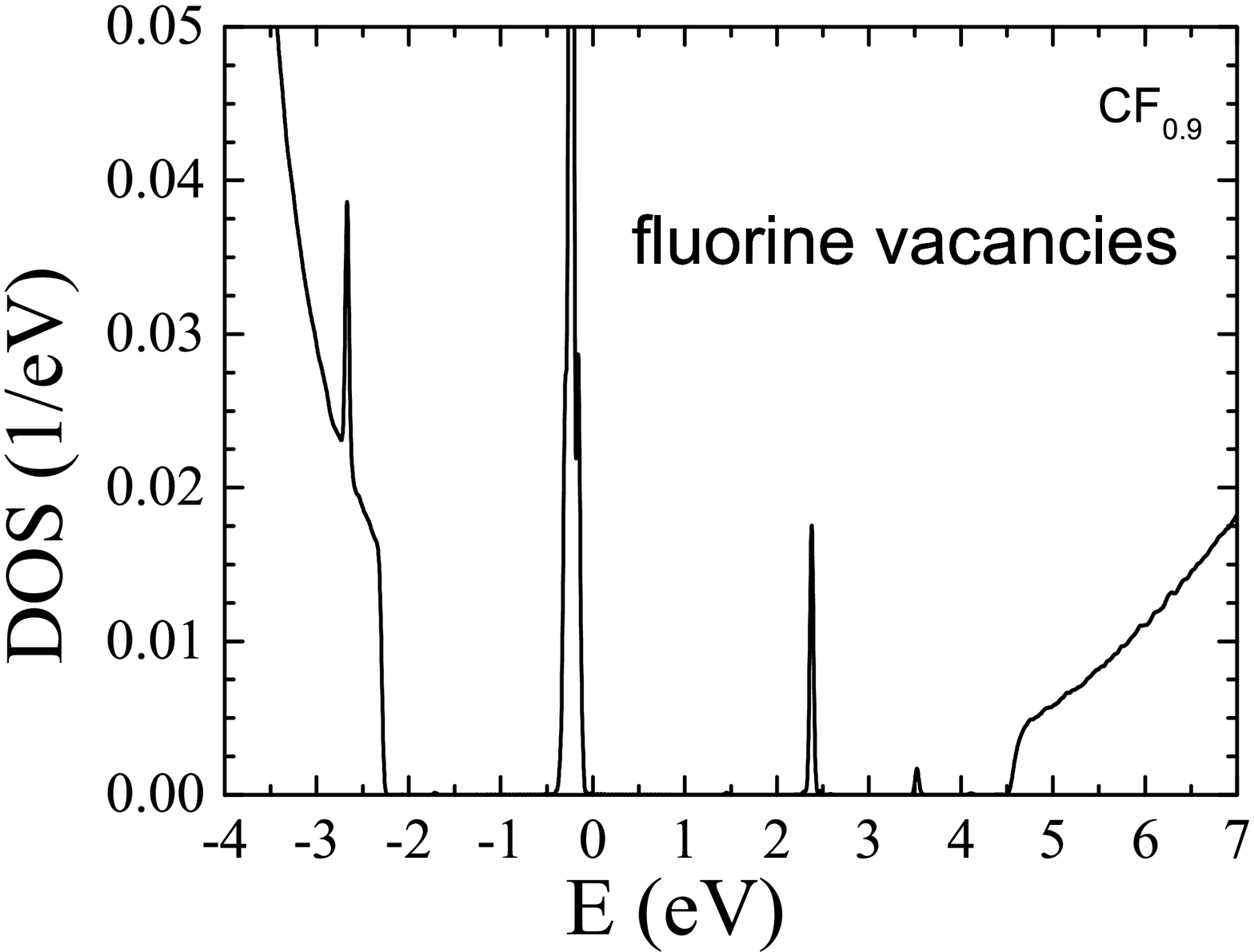}
      \includegraphics[width=5cm]{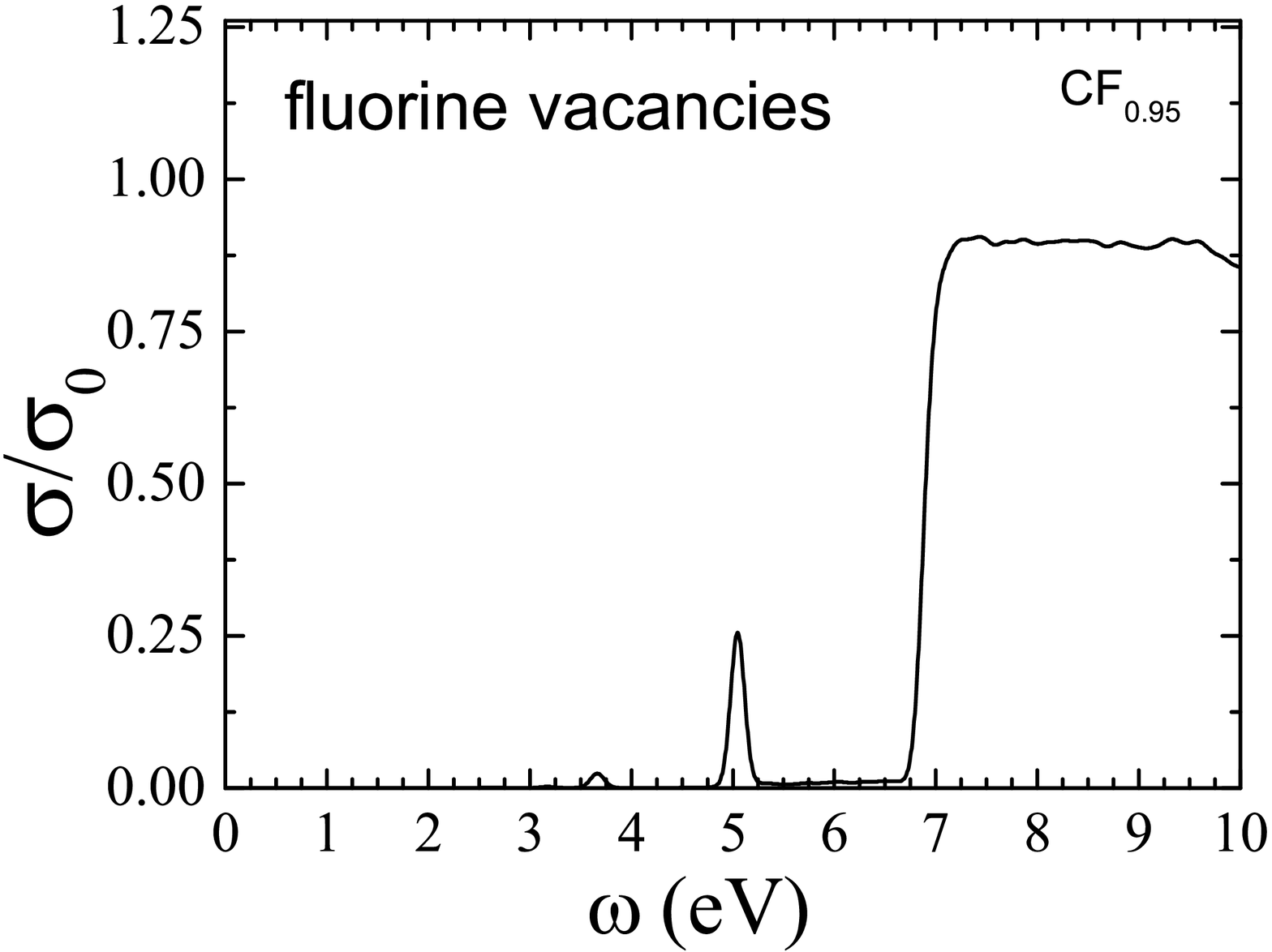}
    }
    \mbox{
      \includegraphics[width=4.5cm]{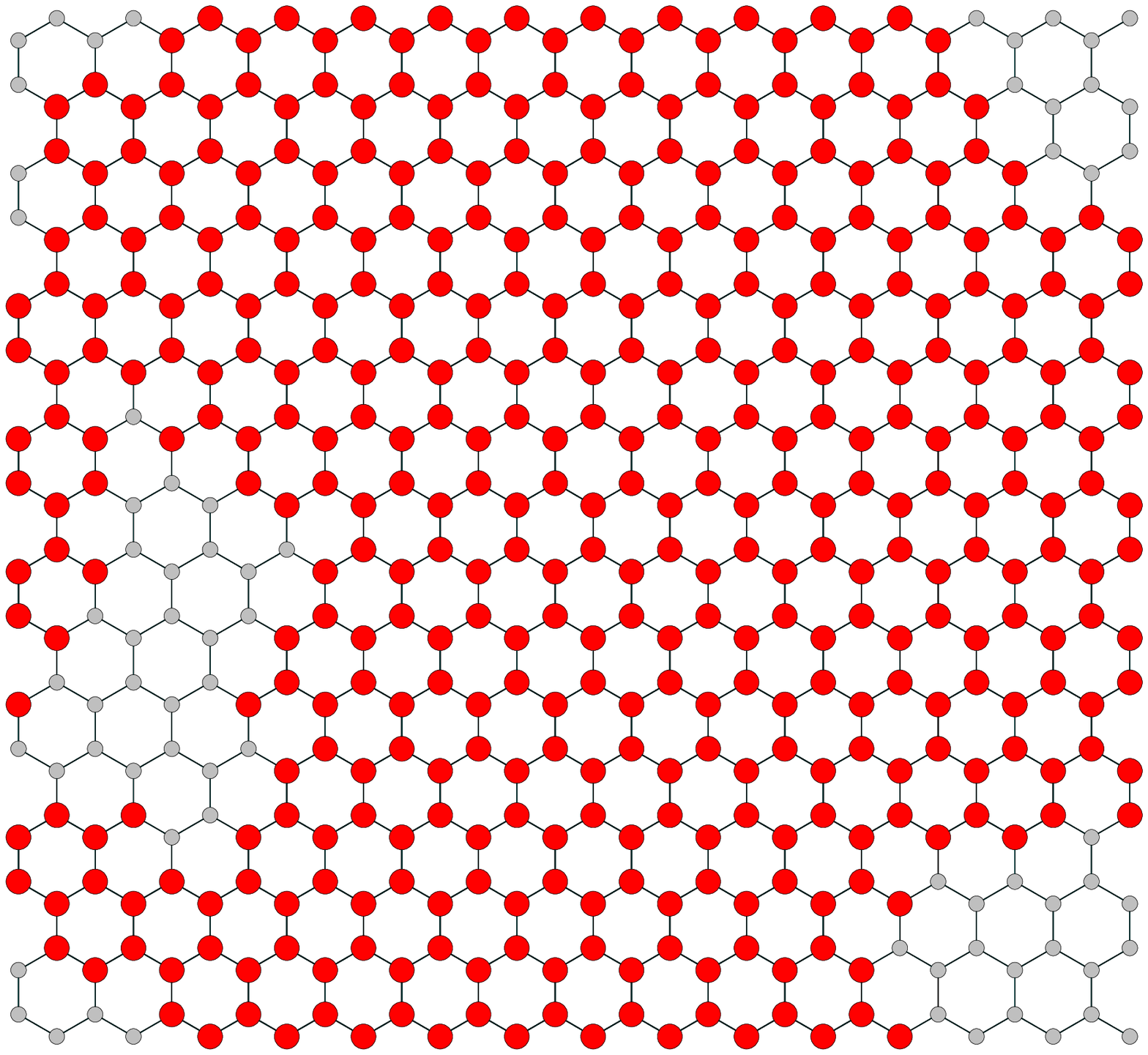}
      \includegraphics[width=5cm]{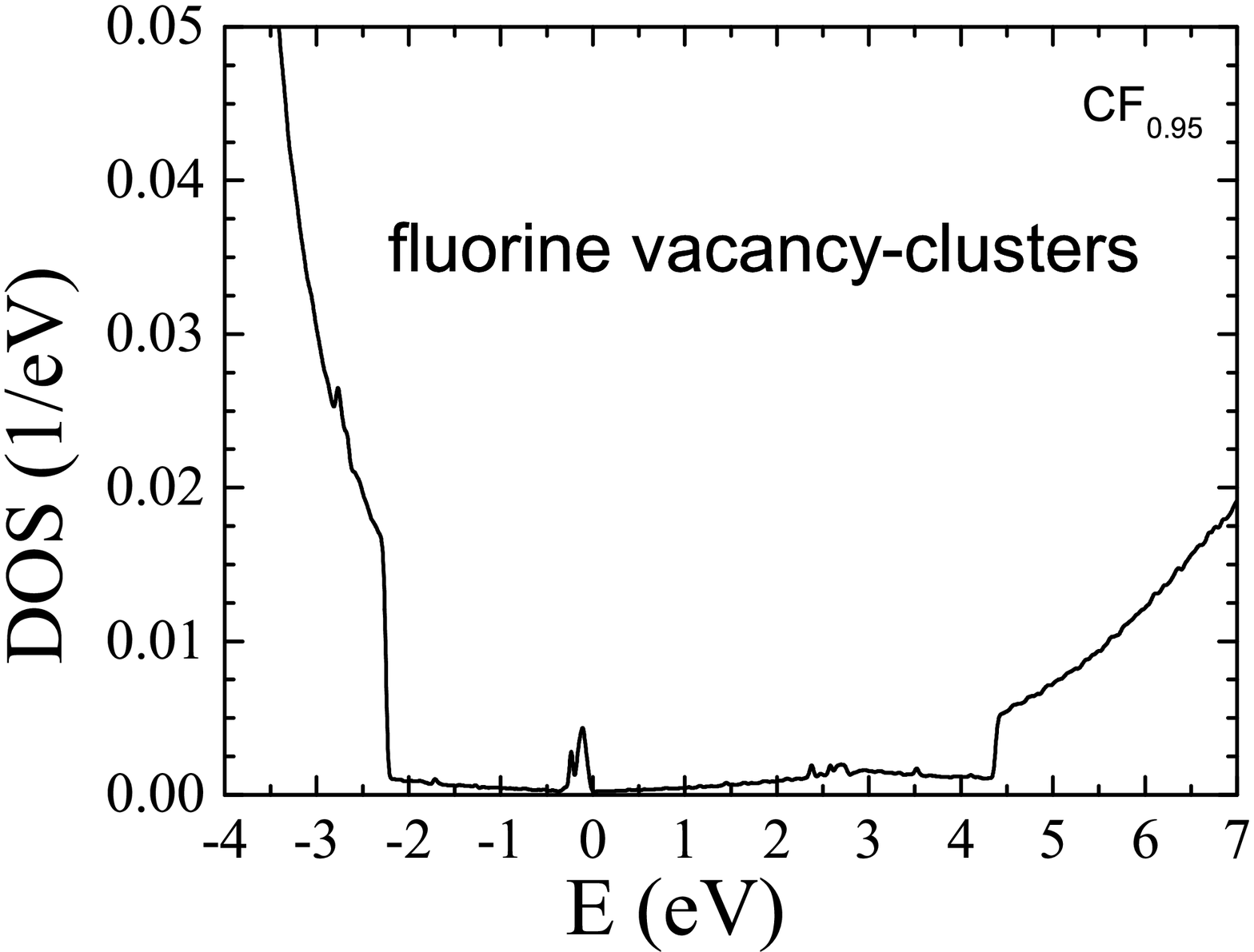}
      \includegraphics[width=5cm]{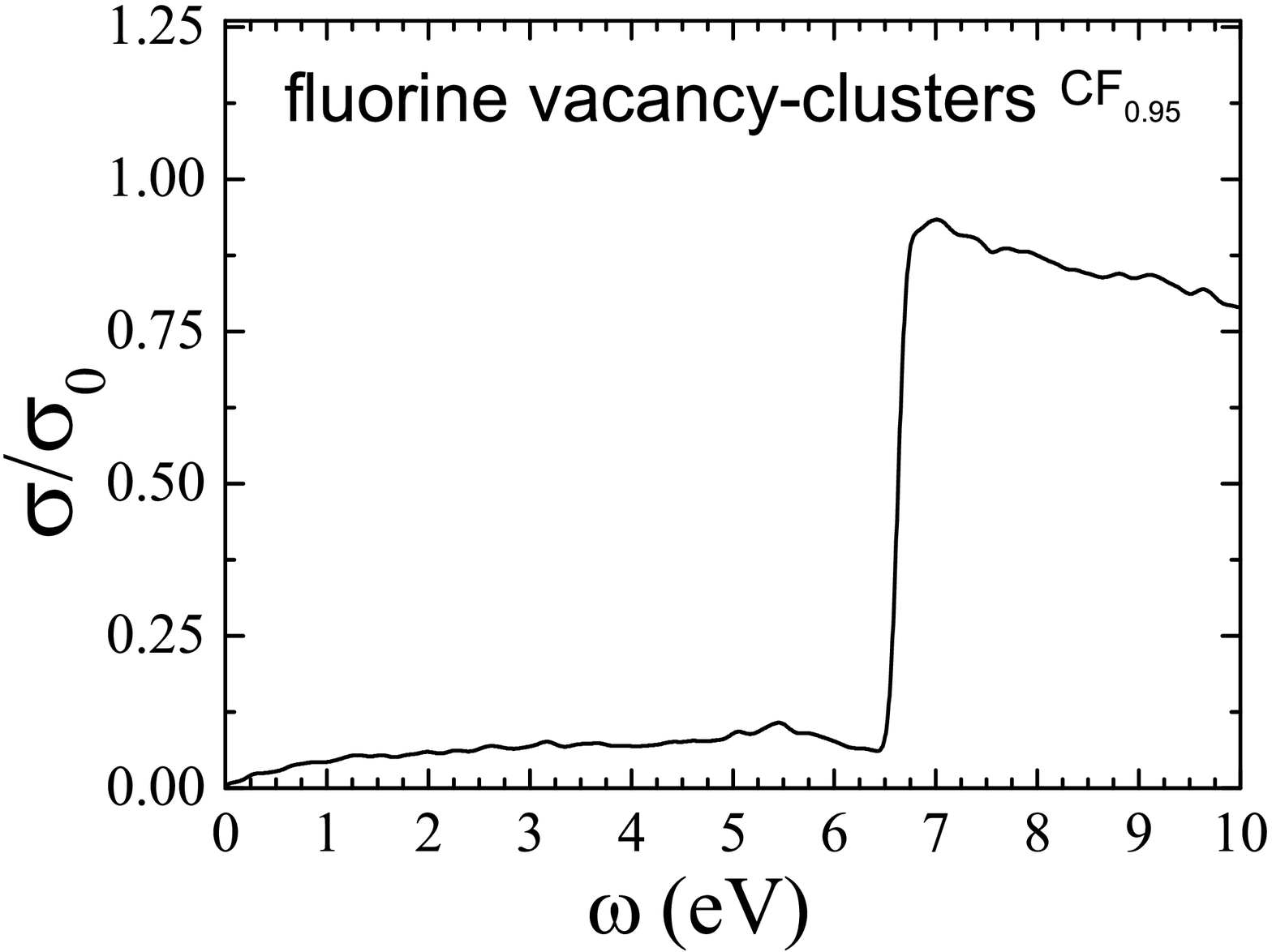}
    }
    \mbox{
      \includegraphics[width=4.5cm]{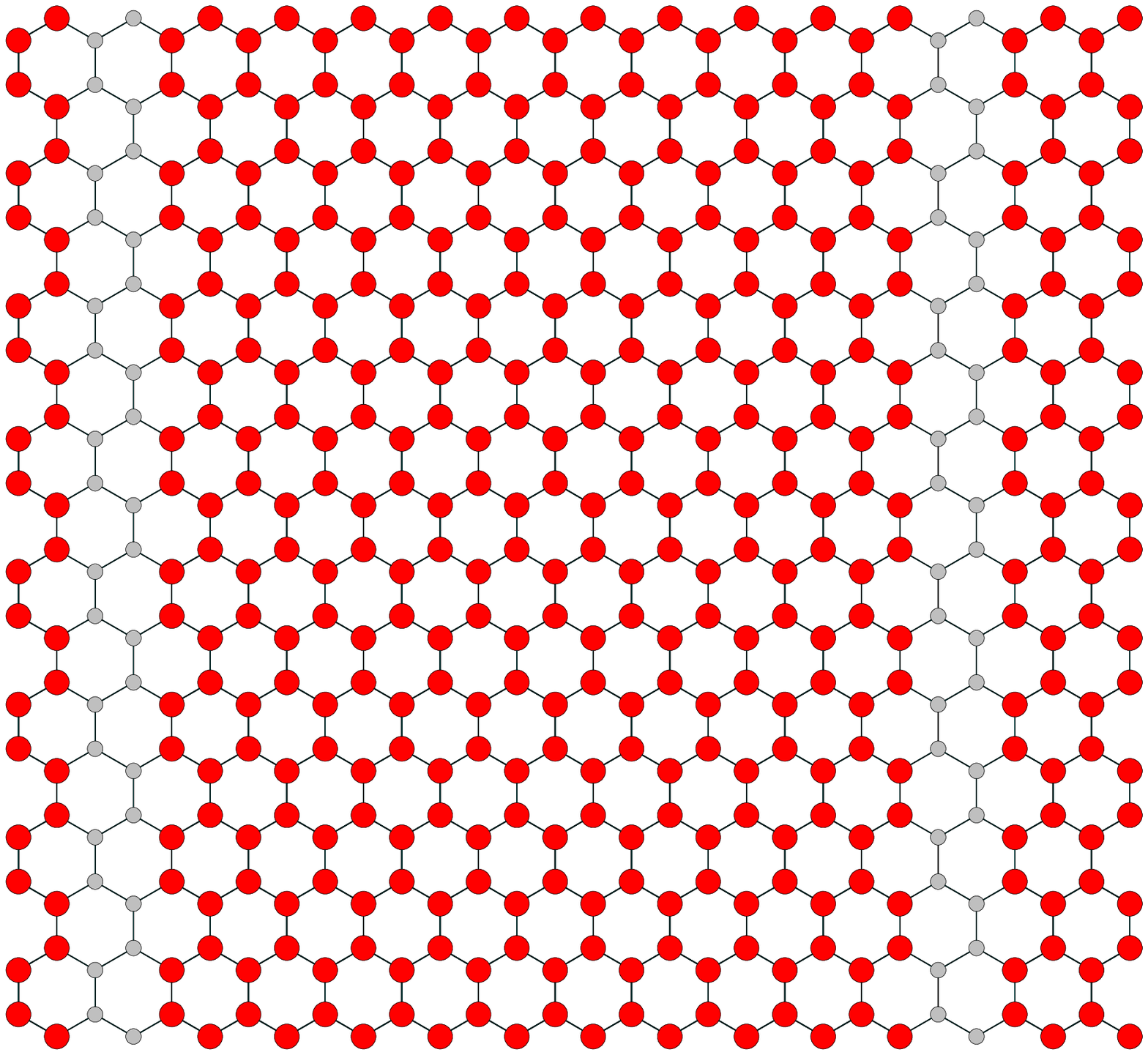}
      \includegraphics[width=5cm]{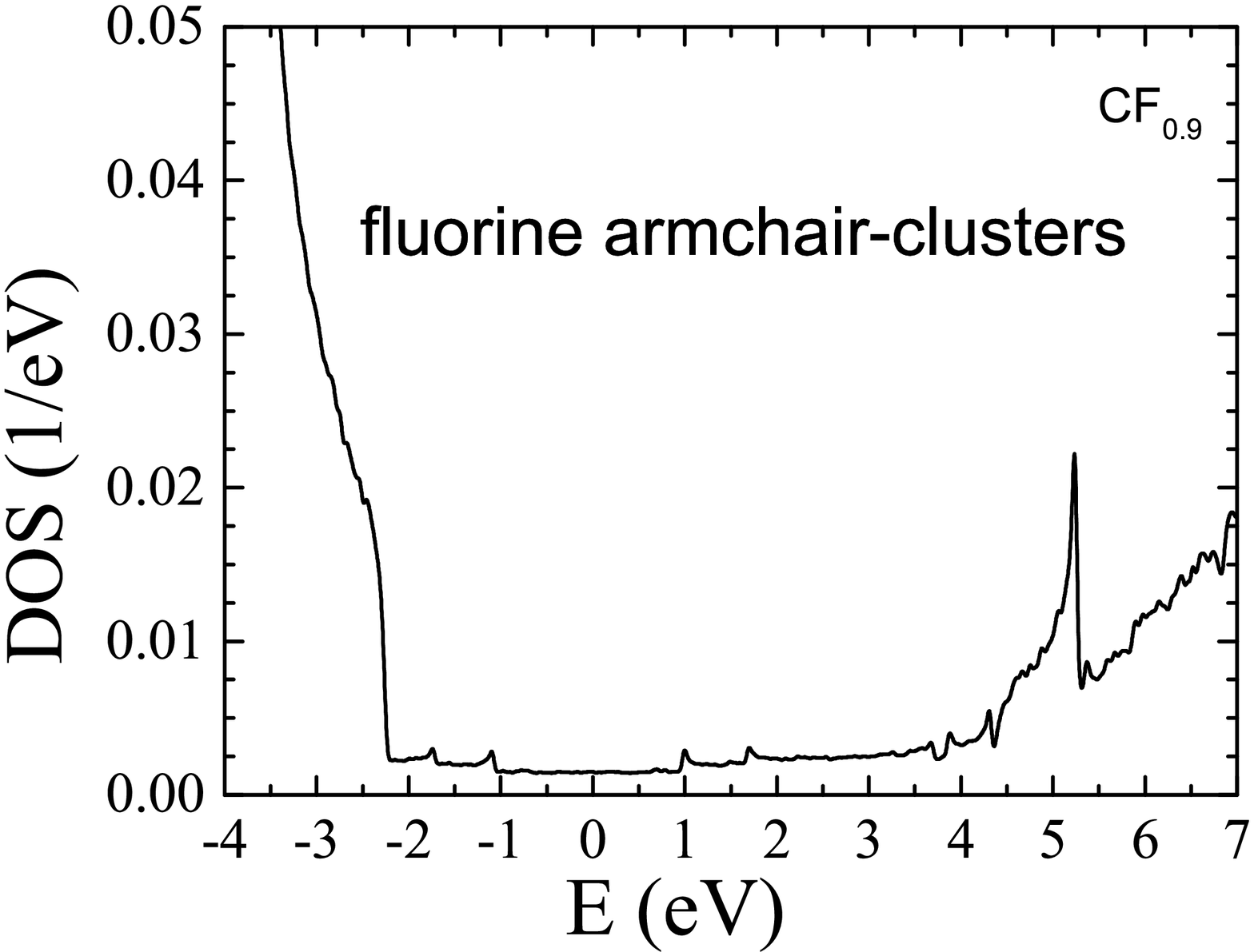}
      \includegraphics[width=5cm]{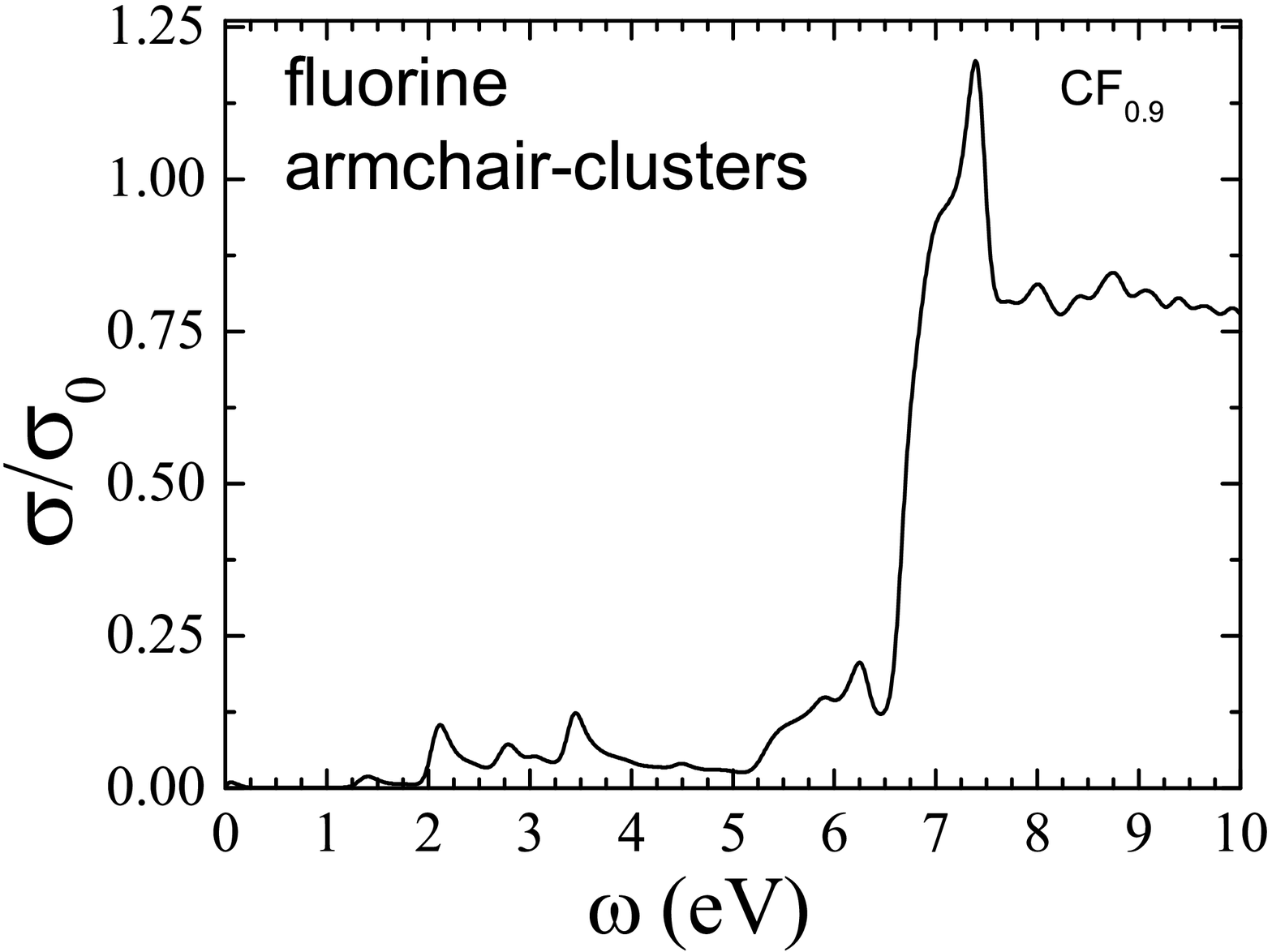}
    }
    \mbox{
      \includegraphics[width=4.5cm]{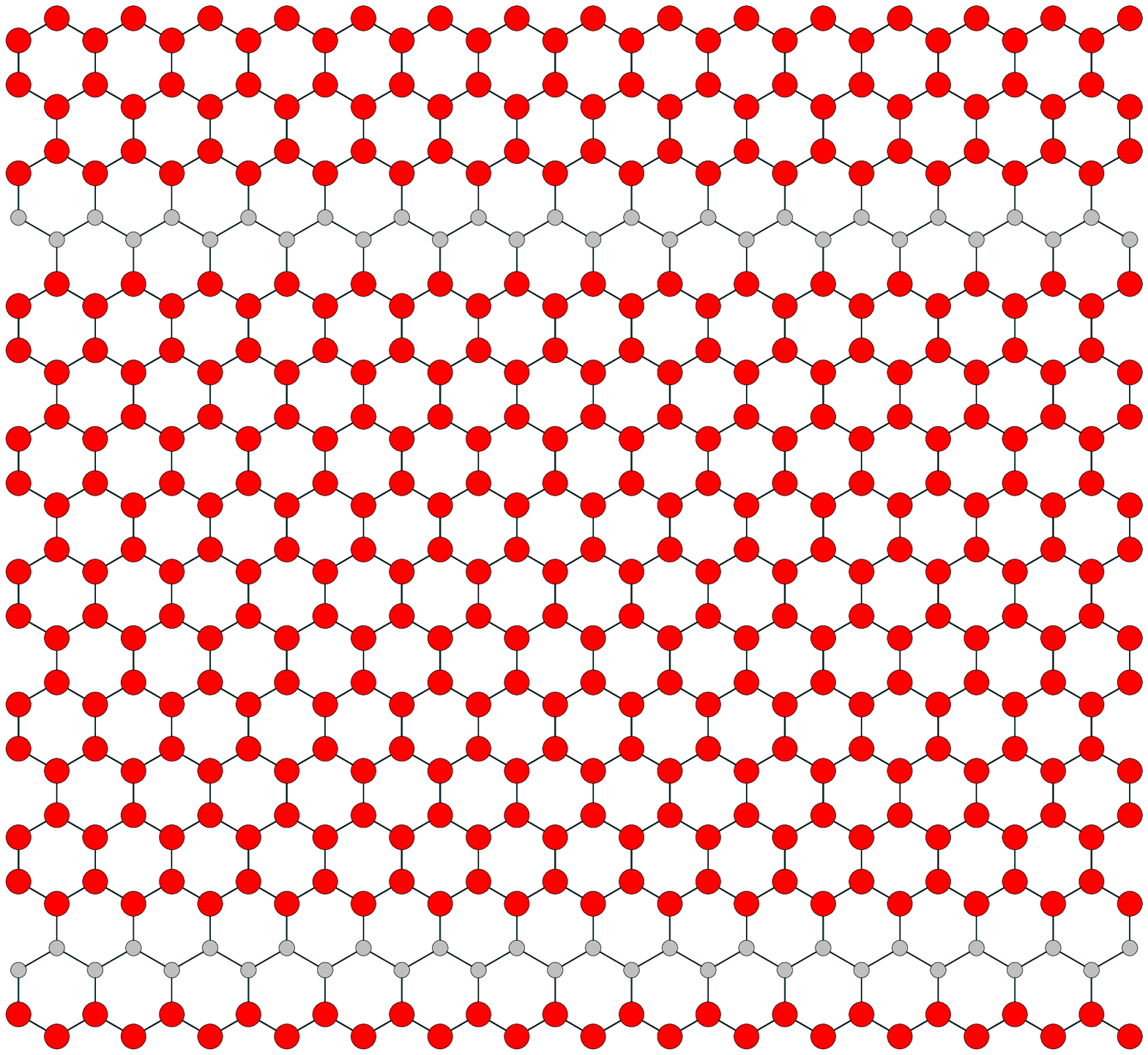}
      \includegraphics[width=5cm]{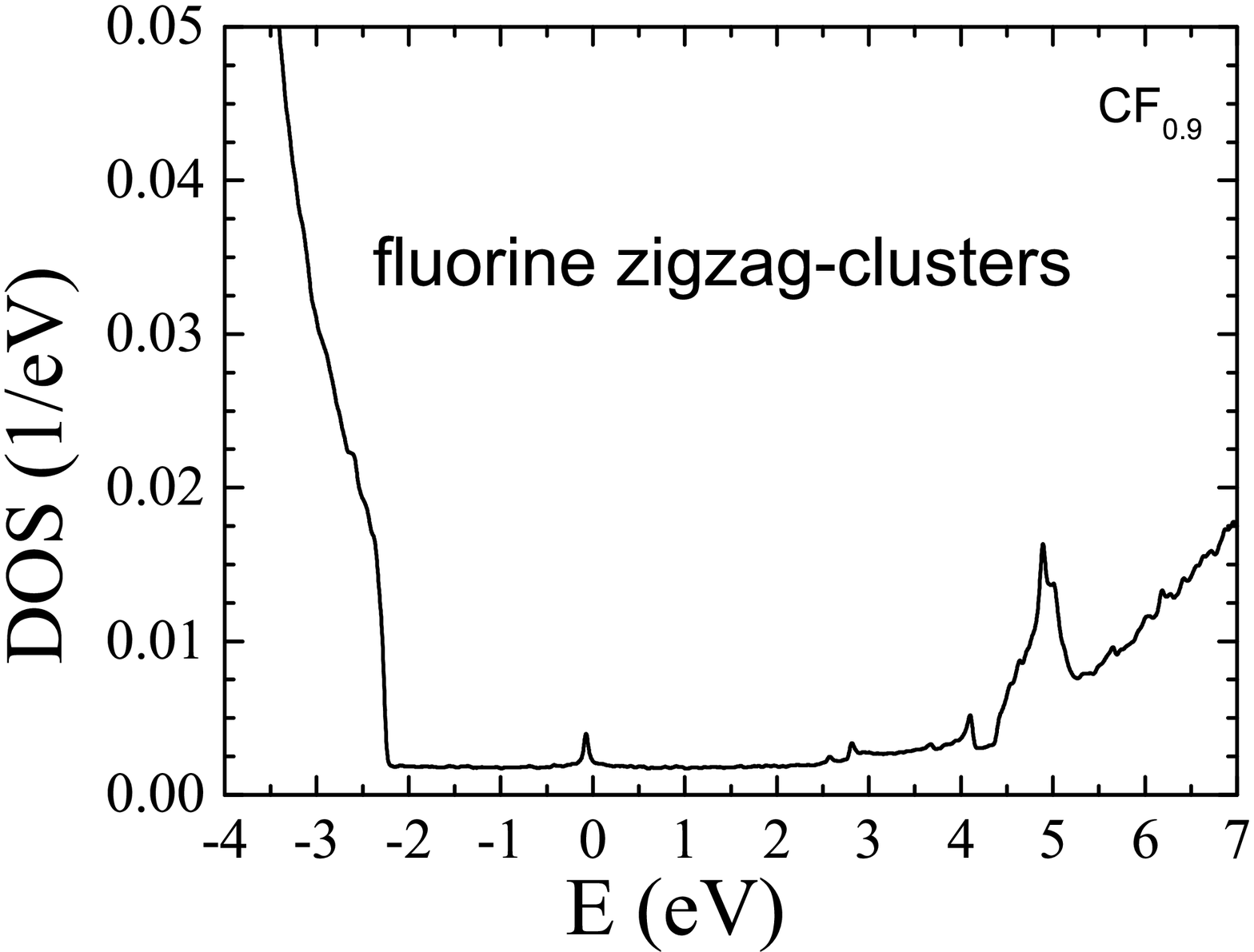}
      \includegraphics[width=5cm]{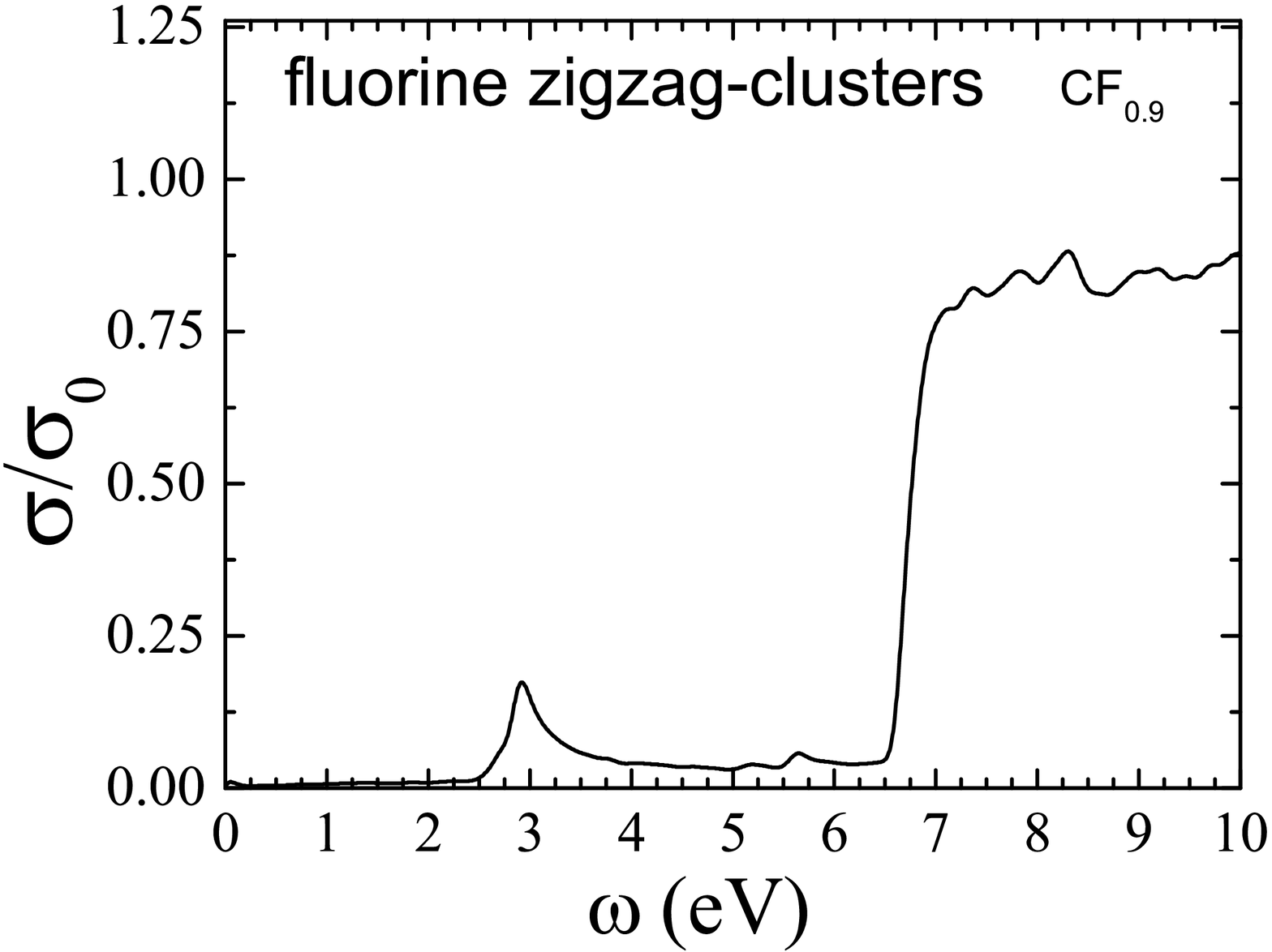}
    }
  \end{center}
  \caption{Left column: Atomic structure with different types of structural disorder.
  The red dots indicate fluorine adatoms. Middle and right columns: density of states and optical conductivity
  of fully or highly fluorinated graphene with different types of structural disorder.
  From top to bottom: Fully or highly fluorinated graphene with (a) randomly
  distributed carbon vacancies; (b) randomly distributed fluorine vacancies;
  (c) randomly distributed fluorine vacancy-clusters; (d) randomly distributed
  fluorine armchair-clusters; (e) randomly distributed fluorine zigzag-clusters.
  For several types of disorder characteristic defect states and features in optical spectra can be identified.}
  \label{Fig:disorder}
\end{figure*}



For fluorine concentrations bigger than F/C$>50\%$, the atomic structures in the paired and unpaired cases become comparable, leading to similar optical in-plane spectra as well (see the results in Fig.~\ref{Fig:ac} (a) and (b)). Thereby individual peaks below $8\,$eV are the most prominent properties of the optical conductivity for these fluorine concentrations. As mentioned above, these peaks are fingerprints of certain atomic structures \cite{YRK10}. 

To investigate these fingerprints in more detail, Fig.~\ref{Fig:disorder} displays the results of fully and highly fluorinated graphene with structural disorder, including a) carbon vacancies (missing of carbon atoms in the graphene membrane); b) fluorine vacancies (missing of fluorine atoms in fluorographene); c) fluorine vacancy-clusters (missing of groups of fluorine atoms); d) fluorine armchair-clusters (or fluorine armchair-vacancy-lines, i.e., the adsorbed fluorine atoms form clusters along armchair lines) and e) fluorine zigzag-clusters (or fluorine zigzag-vacancy-lines, i.e., the adsorbed fluorine atoms form clusters along zigzag lines). 

The common effects due to the presence of structural disorder are defect states (partially) within the electronic band gap. The exact positions of these intragap states are defined by the type of disorder. For example, the defect resonances in the DOS around $E=0.78\,$eV are due to single carbon vacancies, and around $E=-0.17/2.45\,$eV are due to single/paired fluorine vacancies. 

The excitations between these intragap states and the states above or below the band gap lead to narrow or broad peaks in the optical spectrum under $6.3\,$eV. For fluorine vacancies we find a pronounced peak at about $5\,$eV, which has been already discussed above. For fluorine vacancy-clusters, there are many different intragap states due to different structures, forming a continuous background noise within the optical gap.

In the case of full fluorination the optical absorption sets in at the electronic band gap of $6.3\,$eV as obtained in the \G0W0\ approximation, which neglects many body effects like the occurrence of excitons. 
Although our simulations of partially fluorinated graphene show optical excitations below $6.3\,$eV we do not find a reduced optical gap ($\sim 3\,$eV as it has been observed in the experiment) in any of the considered disorder types. Thus we conclude, that the reduction of the optical gap is not due to structural disorder alone.

\begin{figure}[t]
  \begin{center}
    \mbox{
      \includegraphics[width=7cm]{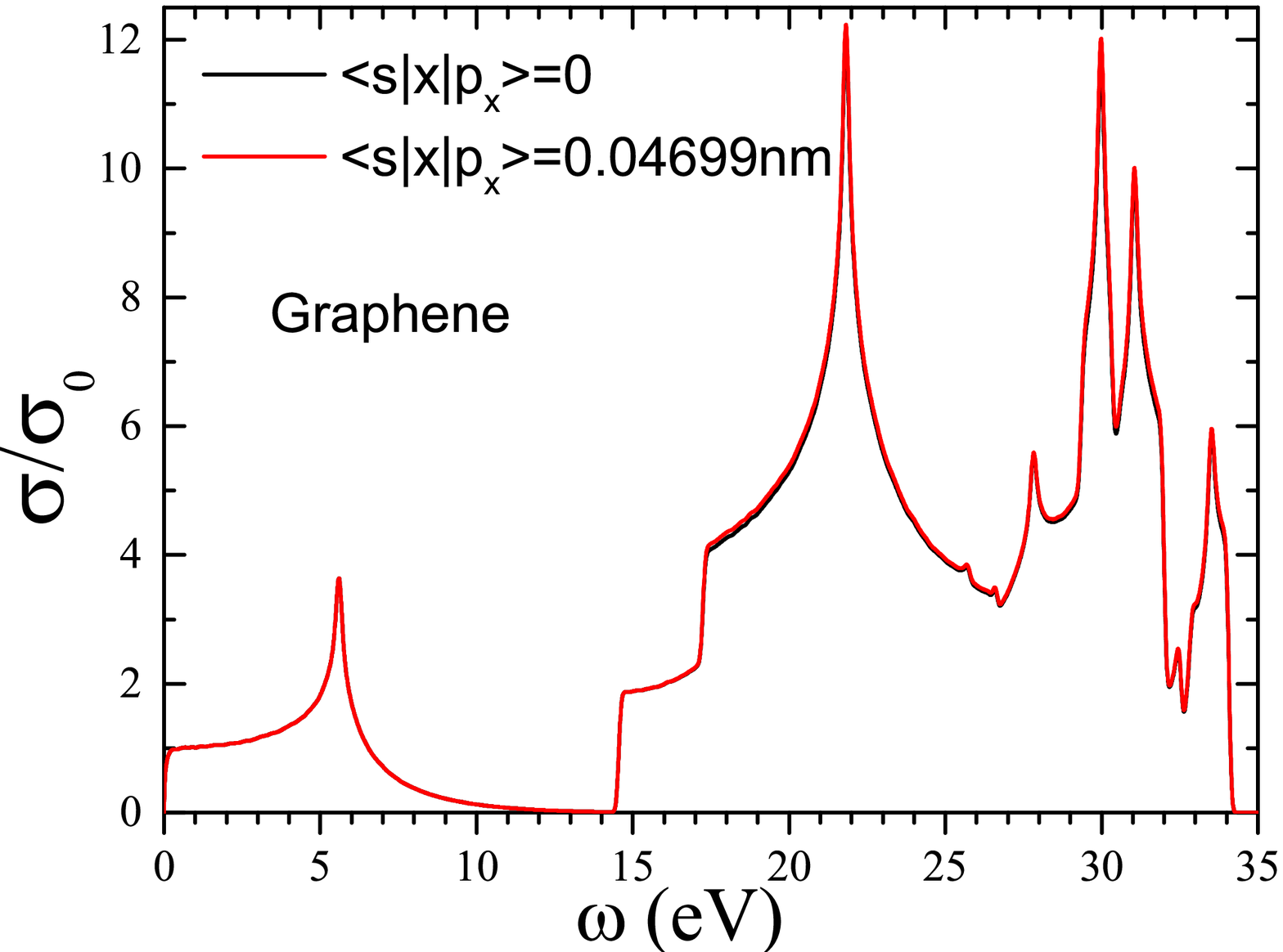}
    }
    \mbox{
      \includegraphics[width=7cm]{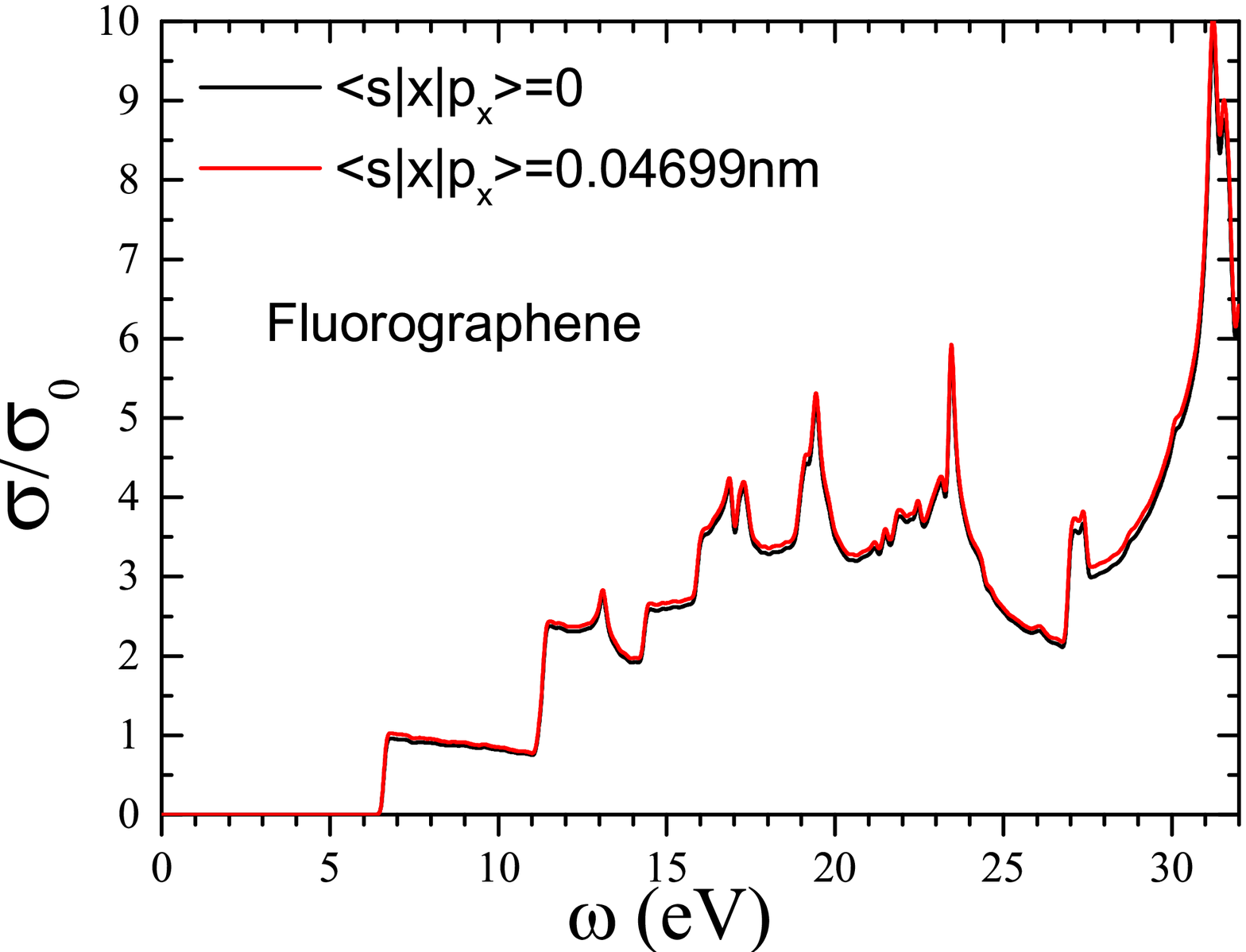}
    }
  \end{center}
  \caption{Comparison of optical conductivity with and without intraatomic dipole
  contribution in graphene (left) and fluorographene (right).}
  \label{Fig:dipole}
\end{figure}

To study the influence of intraatomic dipole contributions to this conclusion we added them to the calculation of the optical conductivity, and found that they are negligible. Fig.~\ref{Fig:dipole} shows the results of graphene and fluorographene with and without intraatomic dipole contributions. The value of the overlap function $\left\langle s|x|p_{x}\right\rangle = \left\langle s|x|p_{y}\right\rangle =0.04699\,$nm are calculated from the overlap of carbon's $s$ and $p_{x}$ ($p_{y}$) wave functions. In general, this dipole contribution slightly increases the value of the optical conductivity. However, a noticeable enhancement of the optical spectra of graphene or fluorographene appears not below energies of $17\,$eV in or $6.3\,$eV, respectively. The dipole contribution does not change the optical spectrum qualitatively and has no effect to the value of the optical band gap in fluorographene. We have also checked that the dipoleterms are also negligible for the out-of-plane optical conductivity (data not shown).

In conclusion, by using a multi-orbital tight-binding model fitted to \abinitio\ calculations, we performed a detailed study of the electronic structure and optical properties of partially and fully fluorinated graphene. For partially fluorinated graphene, the appearance of paired fluorine atoms is found to be more likely than unpaired atoms by matching the simulated optical spectrum to experimental observations. The presence of structural disorder such as carbon vacancies, fluorine vacancies, fluorine vacancy-clusters, fluorine armchair- and zigzag-clusters will introduce defect states within the band gap, leading to characteristic sharp excitations in optical band gap of perfect fluorographene. Both the disorder and excitonic shifts affect the optical spectra on an eV scale and reduce the size of the optical gap. It is thus plausible that their combined effect can reconceile theory and the experimentally observed optical gap. Nevertheless, both mechanisms lead by themselves to sharp resonances below the quasi-particle band gap, which have not been observed experimentally. One would thus have to assume additional broadening of the resonances, e.g. by phonons or further potential fluctuations. Such broadening is ubiquitous in 2d materials \cite{Mak2010,Qiu2013}. Considering the structural change from purely in-plane carbon positions to a buckled structure, we argue that the measurement of polarization rotation of passing 
polarized light through functionalized graphene could be an efficient tool to distinguish between optical effects caused in mainly $sp^3$ as compared to $sp^2$-hybridized regions of the sample. \footnote{See Suppl. Mat.,  which includes Refs. \cite{Ishihara1971,HR00,kresse1994b}}



The support from the European Union Seventh Framework Programme under grant agreement n604391, the Graphene Flagship and from Netherlands National Computing Facilities foundation (NCF) as well as the Norddeutscher Verbund zur F\"orderung des Hoch- und H\"ochstleistungsrechners (HLRN) are acknowledged.

\bibliographystyle{apsrev4-1}
\bibliography{CFBib}

\end{document}